\documentclass[twocolumn,prd,nofootinbib,aps,floats,floatfix,amsmath,amssymb,secnumarabic]
{revtex4}
\usepackage[a4paper,left=1.5cm,right=1.5cm,top=3cm,bottom=3cm]{geometry}

\usepackage{amssymb,amsmath,amsfonts}
\usepackage[dvipsnames]{xcolor}
\usepackage{graphicx}
\usepackage{longtable}
\usepackage{verbatim}
\usepackage{color}

\usepackage{color}
\usepackage{hyperref}
\hypersetup{colorlinks, citecolor=bluscuro, linkcolor=black, urlcolor=bluscuro}
\definecolor{rossos}{cmyk}{0,1,1,0.55}
\definecolor{bluscuro}{rgb}{0.15, 0.2, .85}
\definecolor{bluchiaro}{cmyk}{1,.3,0.,0.1}

\def\bma#1{\mbox{\boldmath{$#1$}}}

\newcommand{\beq}{\begin{equation}}
\newcommand{\eeq}{\end{equation}}
\def\beqa{\begin{eqnarray}}

\def\eeqa{\end{eqnarray}}

\def\lsim{\mathrel{\rlap{\lower4pt\hbox{\hskip0.5pt$\sim$}}
    \raise1pt\hbox{$<$}}}         
\def\gsim{\mathrel{\rlap{\lower4pt\hbox{\hskip0.5pt$\sim$}}
    \raise1pt\hbox{$>$}}}         

\def\pp{{\scriptscriptstyle +}}
\def\mm{{\scriptscriptstyle -}}

\usepackage[normalem]{ulem}

\usepackage{amsmath}
\usepackage{amsfonts}
\usepackage{amssymb}
\usepackage{graphicx, rotating}
\usepackage{epsfig}
\usepackage{latexsym}
\usepackage{graphicx}
\usepackage{color}
\usepackage{amsmath,bm,amssymb}
\usepackage{slashed}
\usepackage{hyperref}
\hypersetup{colorlinks, citecolor=bluscuro, linkcolor=black, urlcolor=bluscuro}
\definecolor{rossos}{cmyk}{0,1,1,0.55}
\definecolor{bluscuro}{rgb}{0.15, 0.2, .85}
\definecolor{bluchiaro}{cmyk}{1,.3,0.,0.1}

\newcommand{\be}{\begin{equation}}
\newcommand{\ee}{\end{equation}}
\newcommand{\bea}{\begin{eqnarray}}
\newcommand{\eea}{\end{eqnarray}}

\def\pp{{\scriptscriptstyle +}}
\def\mm{{\scriptscriptstyle -}}

\newcommand{\arXiv}[2]{\href{http://arxiv.org/pdf/#1}{{\tt [#2/#1]}}}
\newcommand{\arXivold}[1]{\href{http://arxiv.org/pdf/hep-#1}{{\tt [#1]}}}

\def\bma#1{\mbox{\boldmath{$#1$}}}

\begin{document}
\allowdisplaybreaks

\title{\Large\bf\color{black} Tunneling Potentials to Nothing} 
\author{\large J.J. Blanco-Pillado}
\affiliation{ 
Department of Theoretical Physics, University of the Basque Country UPV/EHU}
\affiliation{
EHU Quantum Center, University of the Basque Country, UPV/EHU}
\affiliation{ IKERBASQUE, Basque Foundation for Science, 48011, Bilbao, Spain.}
\author{\large J.R.~Espinosa}
\affiliation{Instituto de F\'{\i}sica Te\'orica, IFT-UAM/CSIC,
C/ Nicol\'as Cabrera 13-15, Campus de Cantoblanco, 28049, Madrid, Spain}
\author{\large J. Huertas}
\affiliation{Instituto de F\'{\i}sica Te\'orica, IFT-UAM/CSIC,
C/ Nicol\'as Cabrera 13-15, Campus de Cantoblanco, 28049, Madrid, Spain}
\author{\large K. Sousa} 
\affiliation{  University of Alcal\'a, Department of Physics and Mathematics,  Pza. San Diego, s/n, 28801, Alcal\'a de Henares (Madrid), Spain.}

\begin{abstract}

The catastrophic decay of a spacetime with compact dimensions, via bubbles of nothing (BoNs), is probably a generic phenomenon. 
BoNs admit a 4-dimensional description as singular Coleman-de Luccia bounces of the size modulus field, stabilized by some potential $V(\phi)$. We apply the tunneling potential approach to this 4d description to provide a very simple picture of BoNs. Using it we identify four different types of BoN, corresponding to different classes of higher dimensional theories. We study the quenching of BoN decays and their interplay with standard vacuum decays.
\end{abstract}

\maketitle

\section{Introduction}
Multiple vacua are common in theories beyond the Standard Model 
and their decay has been widely studied using Euclidean bounces \cite{Coleman,CdL}. In theories with compact extra dimensions, a qualitatively new decay process, mediated by the so-called {\it bubble of nothing} (BoN), was first discussed by Witten \cite{BoN} for the $\mathbb{M}^4\times S^1$ Kaluza-Klein (KK) model. For BoNs in cases with more general internal manifolds and dimensions, see {\it e.g.} \cite{BS,Blanco-Pillado:2010vdp,Blanco-Pillado:2016xvf,Ooguri:2017njy,Dibitetto:2020csn}. 
A BoN describes a hole in spacetime, where the size of a compact dimension vanishes at the surface of the bubble, leaving nothing in the interior. Once nucleated, the BoN expands, ultimately destroying the parent spacetime.

BoNs are also relevant for the Swampland program: the Cobordism Conjecture \cite{CobConj} states that all consistent quantum gravity theories are cobordant between them and thus, they  must admit a cobordism to nothing. BoNs are such configuration, with spacetime ending smoothly on the BoN core. 

BoNs admit an effective $4d$ description as singular 
Euclidean bounces of the modulus field $\phi$ that controls the compactification size \cite{DFG}.
This bottom-up approach, quite useful to study the impact of the potential $V(\phi)$ present in realistic models, has been used  to get some of the necessary conditions on $V(\phi)$ for the existence of BoNs \cite{DGL}. We follow this $4d$ approach but using the  {\it tunneling potential} method \cite{EEg} (sect.~\ref{sec:Vt}). Vacuum decay is described by a tunneling potential, $V_t(\phi)$, that  minimizes a simple action functional in field space. In this language,  BoNs are described by  $V_t$'s which are unbounded in the 
region where the extra dimension disappears. We describe Witten's BoN in the $V_t$ formalism in sect.~\ref{sec:WBoN}. 

This technique allows to efficiently explore possible BoNs. We  identify (sect.~\ref{sec:BotUp}) four types with characteristic asymptotics as $\phi\to\infty$ (the BoN core) corresponding to different $4+d$ origins (depending on the compact geometry and possible presence of a UV defect).
We  study (sect.~\ref{sec:BCs} and \ref{sec:BoNvsOther}) the action and structure of these BoNs contrasting them with other 
 decay channels, like Coleman-De Luccia (CdL) \cite{CdL} or pseudo-bounces  \cite{PS}.
We also identify and study two kinds of BoN quenching. In the first, the action diverges 
(CdL suppression) and the BoN becomes an end-of-the-world brane, while in the second the action remains finite. 
 We summarize  in sect.~\ref{sec:concl}. For further details on our work, see \cite{BPEHS}.

\section{Tunneling Potential Method\label{sec:Vt}}

The tunneling potential method calculates the action for the decay of a false vacuum of $V(\phi)$ at $\phi_\pp$  by finding the (tunneling potential) function $V_t(\phi)$, (going from $\phi_\pp$ to some $\phi_0$ on the basin of the true vacuum at $\phi_-$) that minimizes the functional \cite{EEg}
\be
 S[V_t]=\frac{6\pi^2}{\kappa^2}\int_{\phi_\pp}^{\phi_0}d\phi\ \frac{(D+V_t')^2}{V_t^2 D}\ ,
\label{SVt}
\ee
where $\kappa=1/m_P^2$, with $m_P$ the reduced Planck mass,
we take $\phi_\pp<\phi_0< \phi_\mm$, $x'\equiv dx/d\phi$, and
\be
D^2\equiv V_t'{}^2+6\kappa (V-V_t)V_t\ .
\label{D2}
\ee
When $V_t$ solves its ``equation of motion'' (EoM), 
\be
(4V_t'-3V')V_t' + 6(V-V_t)\left[V_t''+\kappa(3V-2V_t)\right]=0
\ 
\label{EoMVt}
\ee
$S[V_t]$ reproduces the Euclidean bounce result \cite{CdL}.

\begin{figure}
\begin{center}
\includegraphics[width=0.45\textwidth]{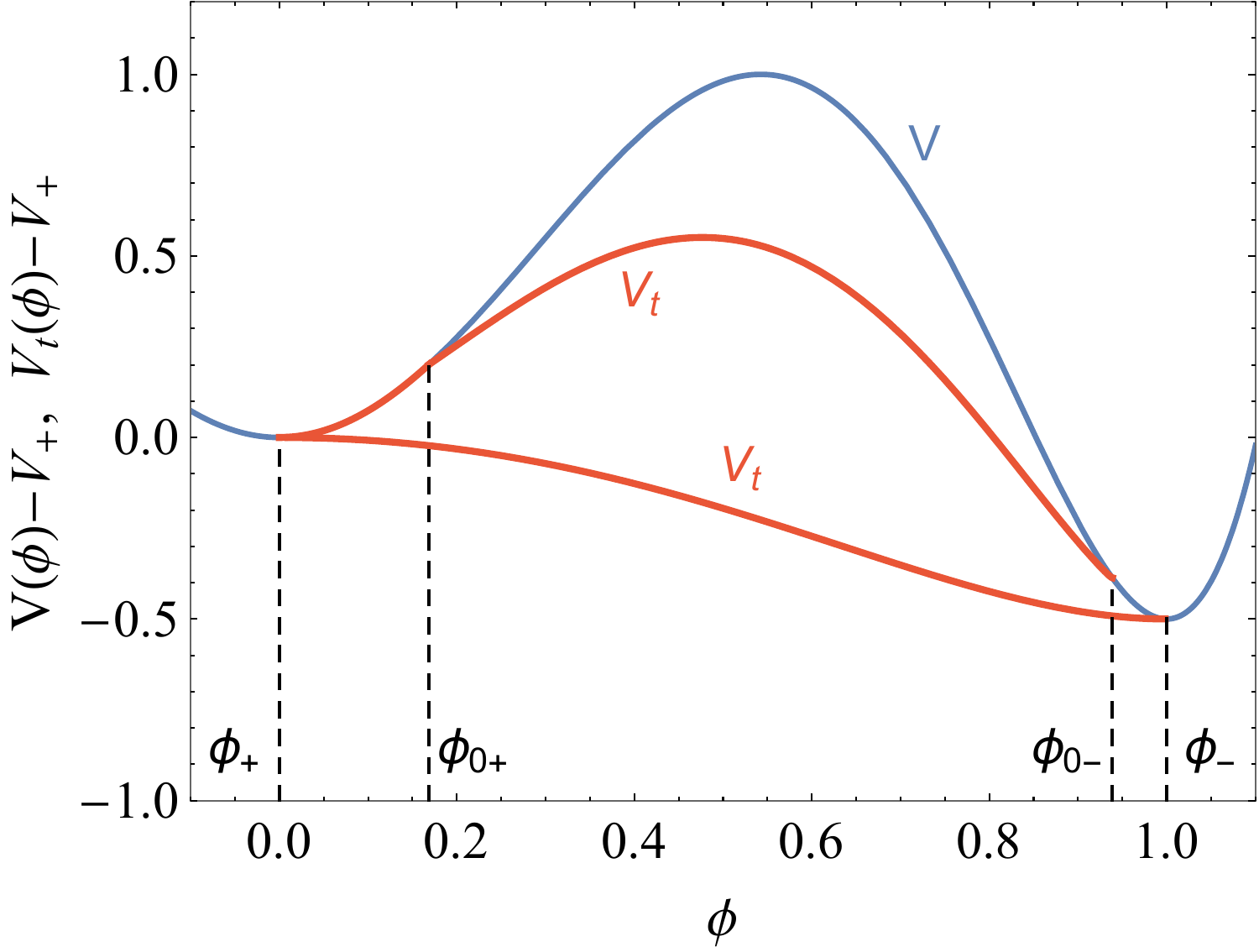}
\end{center}
\caption{Typical shape of tunneling potentials for the decay of AdS/Minkowski (lower curve) or dS (upper) vacua.
\label{fig:Vttypes}
}
\end{figure}

The shape of $V_t$ depends on $V_+\equiv V(\phi_+)$. For $V_+\leq 0$ (Minkowski or AdS false vacua), $V_t$ is monotonic with $V_t,V_t'\leq 0$, see Fig.~\ref{fig:Vttypes}, lower curve.  
Here $\phi_0$ is the core value of the Euclidean bounce, and it is found so as to satisfy the boundary conditions $V_t(\phi_+)= V(\phi_+)$ and $V_t'(\phi_+)=0$ at the false vacuum, and $V_t(\phi_0)= V(\phi_0)$, $V_t'(\phi_0)=3 V'(\phi_0)/4$.

For $V_+>0$ (dS  vacua), $V_t$ is not monotonic and has the shape of the upper curve in Fig.~\ref{fig:Vttypes}, with two parts: A Hawking-Moss (HM) like part  $(\phi_+,\phi_{0+})$, with  $ V_t = V$. A CdL-like part  $(\phi_{0+},\phi_0=\phi_{0-})$ with $V_t<V$. The field values $\phi_{0\pm}\neq \phi_{\pm}$ are found so as to satisfy the boundary conditions
$V_t(\phi_{0\pm}) = V(\phi_{0\pm})$, and $V_t'(\phi_{0\pm})=3V'(\phi_{0\pm})/4$ and coincide with the extreme values of the Euclidean CdL bounce.
If $V_\pp$ grows, the CdL interval shrinks to zero, there is no CdL decay, and the action tends to the HM one \cite{EEg}.

Gravitational quenching of decay, and thus vacuum stabilization, occurs if $D^2>0$ (needed for a real $S[V_t]$) cannot be satisfied for any 
$V_t$. This can happen for Minkowski or AdS vacua if gravitational effects are strong.
Define $\overline{V_t}(\phi)$ as the solution to  $D\equiv 0$  (we set $\kappa=1$ from now on)
\be
\overline{V_t}'= -\sqrt{6 (V-\overline{V_t})(-\overline{V_t})}\ ,
\label{Vtc}
\ee
with $\overline{V_t}(\phi_\pp)=V_\pp$.
To have $D^2>0$, $V_t$ should have slope steeper than $\overline{V_t}$, so that $V_t(\phi)< \overline{V_t}(\phi)$. If, after leaving $\phi_\pp$, $\overline{V_t}$ does not intersect $V$ again, we have quenching. If $\overline{V_t}$ reaches $V$ right at the minimum $\phi_\mm$ (critical case) $V_t=\overline{V_t}$ describes a flat and static domain wall  between false and true vacua, its action is infinite, and gravity also forbids the decay.

In the Euclidean approach, assuming $O(4)$-symmetry, vacuum decay is described by a bounce configuration $\phi(\xi)$, that extremizes the Euclidean action, and a metric function, $\rho(\xi)$, entering the Euclidean metric 
$ds^2= d\xi^2 +\rho(\xi)^2 d\Omega_3^2$.
Here $\xi$ is a radial coordinate and $d\Omega_3^2$ is the line element on a unit three-sphere. 
A dictionary between Euclidean and $V_t$ methods  follows from the key link between both formalisms, $V_t (\phi)= V(\phi) -\dot\phi^2/2$
where $\dot x\equiv dx/d\xi$. 
The  profiles $\phi(\xi)$ and $\rho(\xi)$ can be derived from $V_t$ using the previous link and the Euclidean EoMs \cite{EEg}.

Finally, the $V_t$ approach also describes 
pseudo-bounces \cite{PS} as solutions of (\ref{EoMVt}) with  $V_t'(\phi_0)=0$.
These decay modes are not extremals of the action (they would be if $\phi_0$ were held fixed), and have actions larger than the CdL one. They are relevant when there is no CdL solution  \cite{PS}.

\section{Witten's Bubble of Nothing\label{sec:WBoN}}

The 5d KK spacetime (4d Minkowski $\times S^1$) is unstable against semiclassical decay via the nucleation of a BoN, described by the  instanton metric
\be
ds^2= \frac{dr^2}{1-\frac{\mathcal{R}^2}{r^2}}+r^2d\Omega_3^2 + R_{KK}^2 \left(1-\frac{\mathcal{R}^2}{r^2}\right)d\theta_5^2\ ,
\label{BoNmetric}
\ee
where $R_{KK}$ is the KK radius, $\mathcal{R}$ is the size of the nucleated bubble, $r\in [\mathcal{R},\infty)$, and $\theta_5\in [0,2 \pi)$ parametrises the KK circle. For $r\to\infty$ this metric tends to $\mathbb{M}^4\times S^1$.  
This instanton solution, analytically continued to Lorentzian signature,  describes the tunneling from the homogeneous $\mathbb{M}^4\times S^1$ to a spacetime in which the radius of the 5th dimension shrinks to zero as $r\to \mathcal{R}$ \cite{BoN}.
This BoN ``hole''  at $r=\mathcal{R}$ then expands  and destroys the KK spacetime.   
The decay rate per unit volume  is $\Gamma/V\sim e^{-\Delta S_E}$, with $\Delta S_E=(\pi m_P R_{KK})^2$ the difference between the Euclidean action of the bounce and the KK vacuum. 

The  BoN \eqref{BoNmetric} can be reduced to a 4d description \cite{DFG} integrating the 5th dimension $\theta_5$, and introducing the modulus scalar $\phi$ with
$e^{-2\sqrt{2/3}\,\phi}\equiv 1-R^2_{KK}/r^2$. 
A Weyl rescaling puts the BoN metric 
into CdL form, $ds^2=d\xi^2+\rho(\xi)^2d\Omega_3^2$,
with $d\xi/dr\equiv 1/(1-R^2_{KK}/r^2)^{1/4}$. 
This maps  (\ref{BoNmetric})
into a field profile, $\phi(\xi)$, with the BoN core at $\phi\to\infty$ ($\xi\to 0$) and the KK vacuum at $\phi\to 0$ ($\xi \to \infty$). 
This CdL solution is not of the standard form as the field diverges at 
$\xi=0$. Nevertheless, its Euclidean action 
is finite and equal to Witten's (after including a boundary term of 5d origin). 

Finding the description  
in the $V_t$ approach is straightforward,  using $V_t=V-\dot\phi^2/2$. One gets
\be
V_t(\phi) = -
(6/R^2_{KK})
\sinh^3(\sqrt{2/3}\,\phi)\ ,
\label{WBoN}
\ee 
with $V_t(0)=0\ , \quad V_t(\phi\rightarrow\infty)\sim - e^{\sqrt{6}\phi}$ ,
so that $V_t$ diverges at $\phi\rightarrow \infty$. This is a generic property of the $V_t$'s of BoNs.
Furthermore, the action in the $V_t$ formalism, Eq. (\ref{SVt}), gives the correct result without the need of additional boundary terms. This is true for any other BoN solution in this formalism, see \cite{BPEHS}. 

\section{BoNs with Nonzero Potential
\label{sec:BotUp}}

The modulus field potential, $V(\phi)$, needed to stabilize the extra dimensions, affects the existence and shape of BoNs. In the spirit of \cite{DGL}, we derive the conditions that $V(\phi)$ must satisfy to allow BoN decays. The single function $V_t(\phi)$, on the same footing as $V(\phi)$, captures the key BoN asymptotics in a simple way.  Without assumptions about the origin of $V(\phi)$, we first identify four different types of asymptotics of $V$ and $V_t$ compatible with BoNs. 

Any $V_t$ describing a BoN solves Eq. (\ref{EoMVt}) with standard boundary conditions at the false vacuum $\phi_+$, and $V_t\to-\infty$ at $\phi\to\infty$ (the BoN core).
The four different types of core
asymptotics depending on the value of $\lim_{\phi\to\infty}V/|V_t|$ are listed in Table~\ref{table:types}:

{\bf Type 0}: $\lim_{\phi\to\infty}V/|V_t|=0$. Whether $V$ is positive or negative at $\phi\to\infty$, (\ref{EoMVt}) gives $V_t(\phi\to\infty) \sim V_{tA} e^{\sqrt{6}\phi}$, with $V_{tA} <0$.
$V$ is irrelevant for $\phi\to\infty$ and these BoNs  behave as Witten's BoN. 

{\bf Types ${\bma \pm}$}:  $\lim_{\phi\to\infty}V/|V_t|$ is a constant of sign $\pm$ which labels the type. For $V\sim V_A e^{a \sqrt{6}\phi}$ and $V_t\sim V_{tA} e^{a \sqrt{6}\phi}$ at $\phi\to\infty$, with $a>0, V_{tA}<0$, (\ref{EoMVt}) gives
\be
[V_A+(a^2-1)V_{tA}](3V_A-2V_{tA})=0\ .
\label{asymptcond}
\ee
The first option is $V_{tA}=V_A/(1-a^2)$. For type $-$: $V_A<0, a<1$; for type $+$:  $V_A>0, a>1$. 

{\bf Type ${\bma -^*}$:} The second option to satisfy (\ref{asymptcond}) is  $V_{tA}=3V_A/2$. One needs $V_A<0$, as $V_t< V$. 

Table~\ref{table:types} also shows additional constraints on $a$ obtained by requiring the finiteness of BoN action.

\begin{table}
\begin{center}
\begin{tabular}{ccccc}
Type & 0 & $-$  & $+$ & $-^*$\\
$V_t(\infty)$ & $V_{tA}e^{\sqrt{6}\phi}$ &
$\frac{V_{A}e^{a\sqrt{6}\phi}}{(1-a^2)}$ &
$\frac{V_{A}e^{a\sqrt{6}\phi}}{(1-a^2)}$ &
$\frac{3V_{A}e^{a\sqrt{6}\phi}}{2}$ \\
 Param.& $V_{tA}<0$ & $V_A<0$ & $V_A>0$ & $V_A<0$\\
     Constr.   & $a<1$  & $\frac{1}{\sqrt{3}}<a< 1$ & $a> 1$  &  
        $a>    \frac{1}{\sqrt{3}}$\\
$\beta$ & $\frac13$ & $\frac{1}{3a^2} $ & $\frac{1}{3a^2} $ & 1\\
$\delta$ & $ \frac{(1+\beta)}{\sqrt{2\beta}}$ &  
$ \frac{(1+\beta)}{\sqrt{2\beta}}$ & $ \frac{(1+\beta)}{\sqrt{2\beta}}$ & $ a\sqrt{6}$ \\
   UV &  $S^1$ &   $S^d$ &   Sing. & Sing.
\end{tabular}
\caption{\it For $V(\phi\to\infty)=V_{A}e^{a\sqrt{6}\phi}$ we show, for the four different types of BoN: the asymptotics of $V_t(\phi\to\infty)$; parameter constraints; the exponent $\beta$ in $\rho(\xi\to 0)\sim \xi^\beta$; the exponent $\delta$ in $D(\phi\to\infty)\sim D_\infty e^{\delta \phi}$;  and their possible UV origin. Label "Sing." indicates the need for a defect to avoid a singularity.
\label{table:types}}
\end{center}
\end{table}

$V$ and $V_t$ determine the asymptotics of the Euclidean BoN functions $\phi(\xi)$ and $\rho(\xi)$ at $\xi\to 0$. We get $\rho\simeq c_\rho\xi^\beta$, with $\beta$ as given in Table~\ref{table:types}, and
$\phi \simeq -\frac{1}{a}\sqrt{\frac{2}{3}}\log \left[\xi a\sqrt{3 (V_A-V_{tA})}\right]
$.
For type $0$ BoNs, this holds with $a=1$, $V_A=0$ and agrees with \cite{DGL}.
Thus, the $4d$ instanton is singular, with the leading behaviour near the singularity determined by $V_{tA}$.  

From a $4+d$ BoN geometry, we can integrate over the compact space to get a reduced 4d metric and 
a modulus field [with  potential $V(\phi)$] that tracks the size of the extra dimensions. 
This gives a 4d picture of the BoN as a singular CdL bounce $\phi(\xi)$ \cite{DFG,DGL}, or as a divergent tunneling potential, $V_t(\phi)$. Via such top-down approach we explore the $4+d$ origin of the parameters in the  BoNs found above. 

Consider first a BoN with a $d$-dimensional sphere, $S^d$, of radius $R_{KK}$, as compact space.  Imposing the smoothness of  the $4+d$ BoN solution at $r\to 0$, and reducing to $4d$, we obtain the $\xi\to 0$ scaling
\be
\phi \simeq -\sqrt{\frac{2d}{(d+2)}}\ \log\xi_d\ ,\quad
\rho \simeq {\cal R}\ \xi_d^{d/(d+2)}, 
\label{phirhotop}
\ee
where $\xi_d\equiv (d+2)\xi/(2R_{KK})$,
which agrees with  \cite{DGL}.  
Comparing with the scalings found above using $V_t$, we get  
$a=\sqrt{(d+2)/(3d)}$, $\beta =d/(d+2)$
and
\be
D_\infty =  \frac{3}{R_{KK}{\cal R}}\sqrt{\frac{d(d+2)}{2}}
\ .
\label{UVresults}
\ee
where $D(\phi\to \infty) \sim D_\infty e^{\delta \phi}$, with $\delta$ given in Table~\ref{table:types}.
$V_A$ and $V_{tA}$ are also determined by \eqref{phirhotop}, giving 
\be
V\simeq \frac{-d(d-1)}{2 R_{KK}^2}e^{\sqrt{\frac{2  (d+2)}{d}}\phi},\, V_t \simeq \frac{-3d^2}{4 R_{KK}^2}e^{\sqrt{\frac{2(d+2) }{d}}\phi},
\label{VUVminus}
\ee
 at $\phi\to \infty$. Thus, the smoothness condition imposes $V$ to be of the form one gets in the 4d reduced action from the curvature ${\cal R}_d$ of the compact space
\be
\delta V(\phi) = -\frac{{\cal R}_d}{2}e^{\sqrt{\frac{2(d+2)}{d}}\ \phi}\ .
\ee 
Type 0 BoNs are realized for $d=1$, and type $-$ for $d>1$ [as $1/\sqrt{3}<a=\sqrt{(d+2)/(3d)}<1$]. 

Other well known sources of moduli potentials (see e.g. \cite{DGL}) give 
\be
\delta V(\phi) = \Lambda_{4+d}e^{\sqrt{\frac{2 d}{d+2}}\ \phi}+\frac{Q^2}{2 g^2 \mathcal{V}_{(d)}} e^{3\sqrt{\frac{2d}{d+2}}\ \phi}\ .
\label{VccVflux}
\ee 
The first term comes from a $4+d$ cosmological constant, $\Lambda_{4+d}$. 
If this is the dominant term in $V$, then the $a$ parameter of our $V_t$ description (see table~\ref{table:types}) would be
$1/3\leq  a=\sqrt{d/[3(d+2)]}<1/\sqrt{3}$,  which is of type 0. The second term comes from a $d$-form flux on the compact space,   $\int_{S^{d}} F_{d} =Q$ (with $g$ the gauge coupling and $\mathcal{V}_{(d)}$  the volume of the $d-$sphere). This contribution gives $1\leq  a=\sqrt{3d/(d+2)}<\sqrt{3}$: the scaling of type + cases (provided $d>1$). 

However, for $S^d$ compactifications, the flux contribution to $V$ cannot dominate at $\phi\to \infty$ limit, 
as the regularity conditions require $V$ as in  \eqref{VUVminus}
. Nevertheless, scalar fields present besides the modulus $\phi$ can modify the  potential probed asymptotically by the BoN. Such an example for a BoN in a flux compactification model is given in \cite{BPEHS}. There the naive type $+$ behaviour of the flux contribution is tamed by the presence of a smooth source that effectively transforms the solution in a type $0$ BoN at its core. See \cite{BS} for a higher dimensional realization of this effect.

More exotic types of solutions leading to singular BoNs  (like types $-^*$ and $+$) can also be realized. The BoN singularity signals the need of a brane, or another UV object, whose properties (tension and charge) could be inferred from the behaviour of the solution in the limit  $\phi\to\infty$, see \cite{CobConj,Horowitz:2007pr,Bomans:2021ara,Hebecker}.

\section{Low Field Shooting. BoN Quench \label{sec:BCs}}

For the numerical exploration of vacuum decay solutions, instead of starting at large field values using the overshoot/undershoot method as in \cite{DGL}, we solve the EoM for $V_t$ starting at low field values.  
Our solutions never under/overshoot but are always on target: all starting boundary conditions correspond to a solution, be it a BoN, a CdL or a pseudo-bounce. 

As $V_t$ is a solution of the second order differential equation (\ref{EoMVt}), it depends on two integration constants, {\it e.g.} $V_t$ and $V_t'$ at some field value. For dS vacua, we can solve for $V_t$ starting from the initial point of the CdL range of $V_t$, $\phi_i\neq\phi_\pp$,  with $V_t(\phi_i)=V(\phi_i)$ and $V_t'(\phi_i)=3V'(\phi_i)/4$. For Minkowski or AdS   vacua, we start at $\phi_\pp$
with $V_t(\phi_\pp)=V(\phi_\pp)$ but $V_t'(\phi_\pp)=0$ does not fix completely the solution as $\phi_\pp$ is an accumulation point of an infinite family of solutions, and one needs to impose an additional condition 
to select a particular one, see below.

\begin{figure}
\begin{center}
\includegraphics[width=0.45\textwidth]{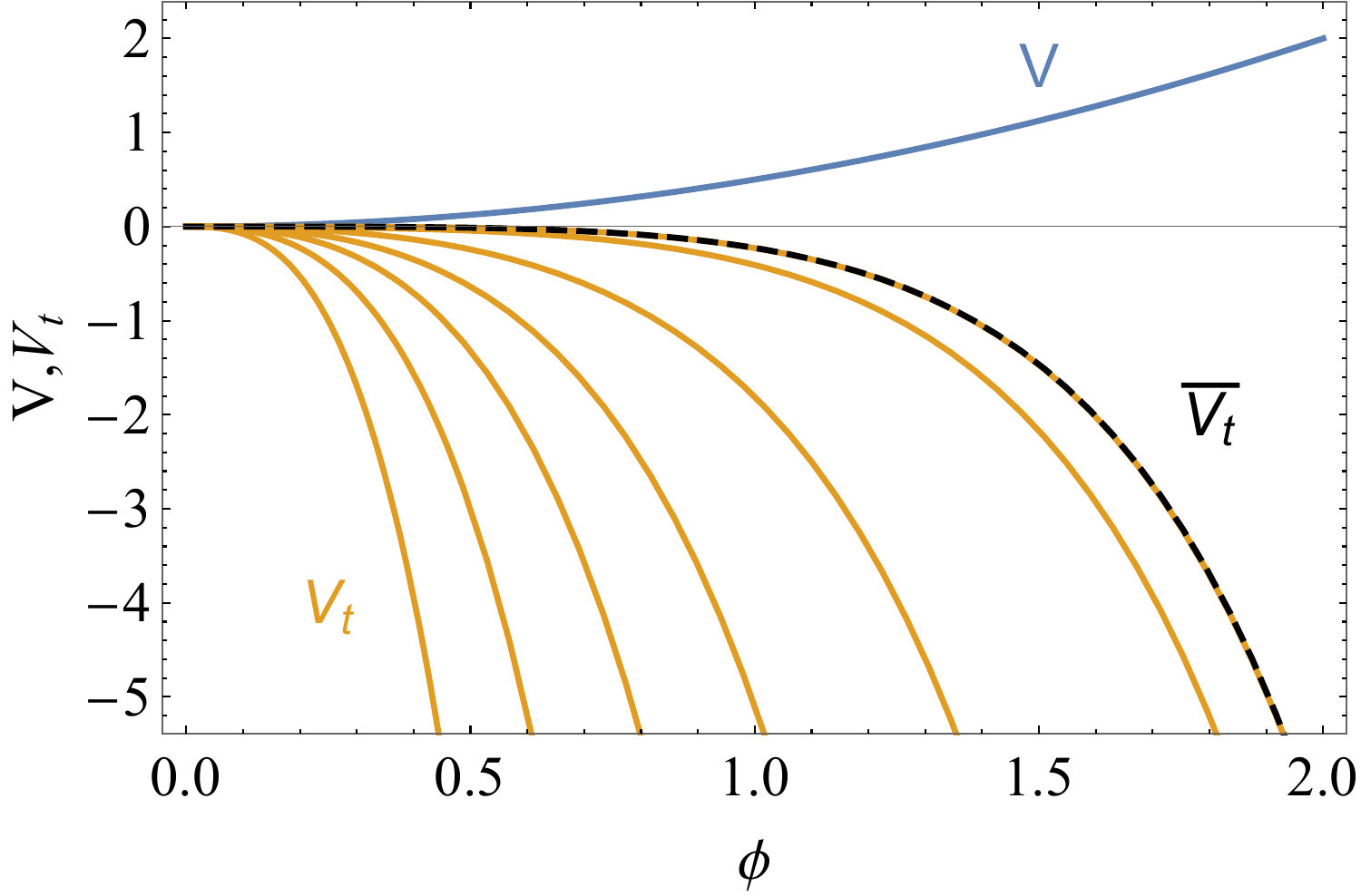}
\includegraphics[width=0.45\textwidth]{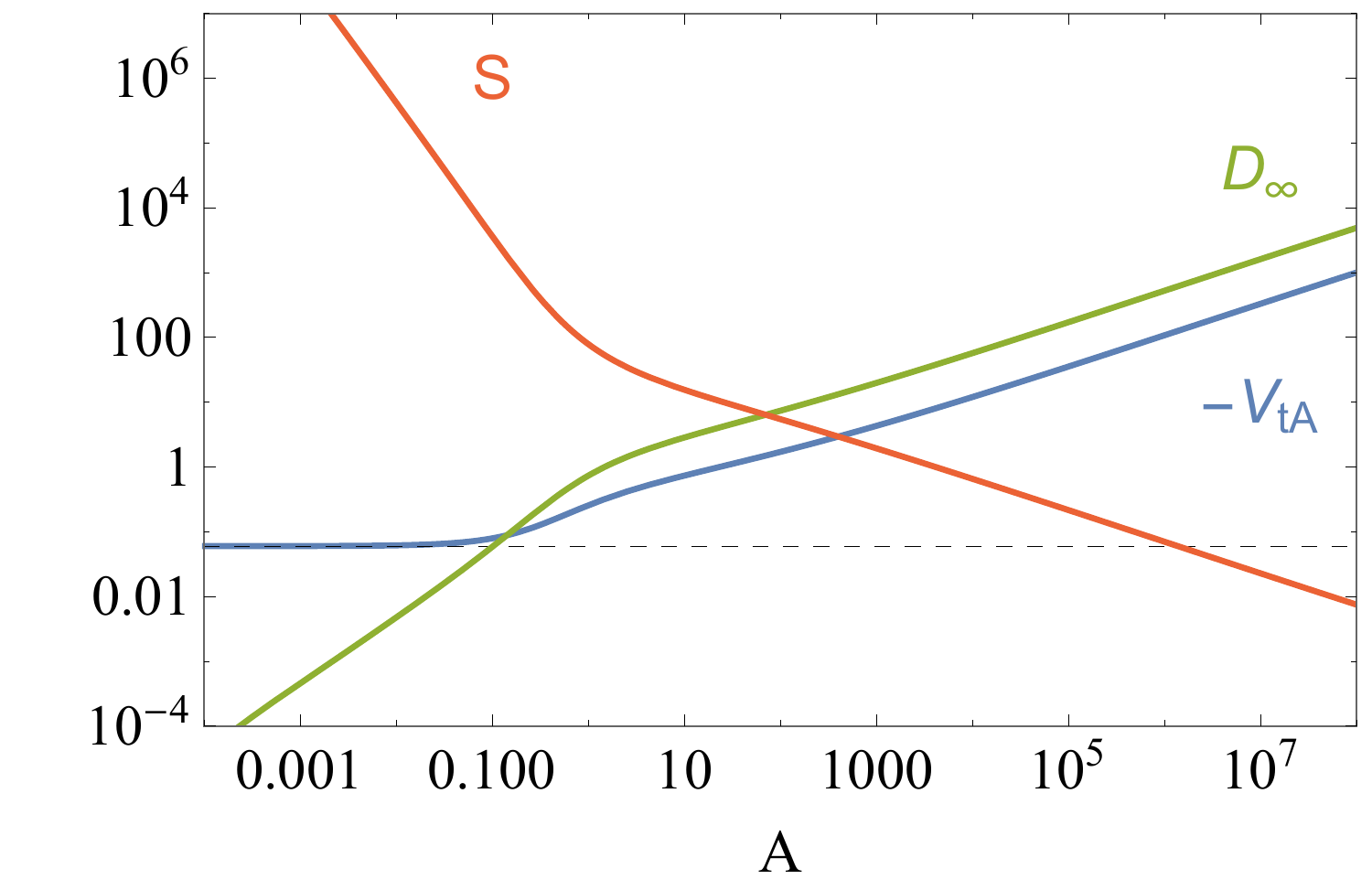}
\end{center}
\caption{Top: Potential $V=\phi^2/2$ and tunneling potentials $V_t(A;\phi)$ (bounded by $\overline{V_t}$, dashed line). Bottom: 
Tunneling action $S$ and prefactors $V_{tA}$ and  $D_\infty$. 
\label{fig:BC0Mink}
}
\end{figure}

There is an interesting interplay between the boundary conditions satisfied by $V_t(\phi)$ at both ends of the field interval in which it is defined. 
In order to illustrate this we use the simple type 0 potential 
$V(\phi)=m^2\phi^2/2$.
The low-field expansion of $V_t$ is
$V_t(A;\phi) \simeq -m^2\phi^2/[1/A-(1/3)\log(A\phi^2)]$,
with  $A>0$ a free parameter. 
(For AdS  vacua the behaviour is similar, with a low field expansion for $V_t$ with a different parameter \cite{BPEHS}.)
We find an infinite family of  solutions, $V_t(A;\phi)$, describing BoN decays, see figure~\ref{fig:BC0Mink}, top plot. For $A\to 0$ we reach the critical  $\overline{V_t}(\phi)$ (black dashed line) 
which has $D=0$ and is an upper limit on allowed $V_t$'s (with $D^2>0$).

The asymptotics of the $V_t$ solutions is of type 0, $V_t\sim V_{tA}(A) e^{\sqrt{6}\phi}$ and $D\sim D_\infty(A) e^{\sqrt{8/3}\phi}$.  (For these BoNs the two integration constants in the large field regime can be  chosen to be $V_{tA}$ and $D_\infty$). The functions $V_{tA}(A)$ and $D_\infty(A)$ depend on $V(\phi)$ and are given for our case in the bottom plot of figure~\ref{fig:BC0Mink}. Interestingly, $-V_{tA}$ is bounded below by the $V_{tA}$ prefactor of $\overline{V_t}\sim \overline{V}_{tA}e^{\sqrt{6}\phi}$ (dashed line). 

As shown in sect.~\ref{sec:WBoN}, a given $4+d$ theory with fixed $R_{KK}$ 
determines $V_{tA}$ via  \eqref{VUVminus} (with $d=1$), thus selecting one member of the family of $V_t$'s.
When $-V_{tA}(A)$ is bounded below,  $-V_{tA}(A)\ge -V_{tA*}$ (as  in fig. \ref{fig:BC0Mink}), BoN decay is allowed provided
\be
R_{KK}^2=\frac{3}{4 (-V_{tA})}\le \frac{3}{4 (-V_{tA*})}\ ,
\label{eq:dyConst1}
\ee
and forbidden otherwise. The critical case  $V_{tA*}$ corresponds to the limit $A\to 0$, for which  $D_\infty\to 0$, $S\to\infty$ and 
${\cal R}\to \infty$ [see \eqref{UVresults}], as expected for an infinite and static BoN: an End-of-the-World brane  \cite{DynCo}. The situation is similar for the AdS case. 

This dynamical obstruction to BoN decay \cite{Blanco-Pillado:2016xvf,GMSV} is similar to the CdL quenching of the standard decay of  Minkowski or AdS false vacua, with the critical case corresponding to a domain wall of infinite action, see Sect.~\ref{sec:Vt}. Although already \cite{BoN} discussed possible topological obstructions to BoN decays, the Cobordism Conjecture \cite{CobConj} 
removes such obstructions. In that case, the only protection of a compactification against BoN formation must be dynamical \cite{Blanco-Pillado:2016xvf,GMSV}.

For the dS case, an expansion of $V_t$ near $\phi_i$ is used and 
one can take $\phi_i$ as the free parameter for a family of $V_t$ solutions. 
For regular CdL decay, we get a family of pseudo-bounces ending at the proper CdL \cite{PS}. 
For BoNs, we get a type 0 family with asymptotics $V_t\sim V_{tA}(\phi_i) e^{\sqrt{6}\phi}$. 
However, now there is no bound on $V_{tA}(\phi_i)$ (as there is no critical $\overline{V_t}$) and thus no dynamical constraint on BoN decay (see \cite{BPEHS}). 

\section{BoNs vs. Other Decay Channels\label{sec:BoNvsOther}}

To illustrate the interplay of BoNs with standard decay channels (CdL decay and pseudo-bounces) let us consider the  potential  
\be
V(\phi)=V_\pp+\frac12 m^2\phi^2-\lambda \phi^4+\lambda_6\phi^6\ ,
\label{Vcex}
\ee
which admits examples of type 0 BoNs.  For numerics  we take $m=1$, $\lambda=17/4$ and $\lambda_6=8/3$. The potential has a false vacuum at $\phi_\pp=0$, separated  from the true vacuum at $\phi_\mm=1$ by a shallow barrier that peaks at $\phi_B=0.25$.  (Further examples, including dS cases with HM but no CdL decay, type $-$ BoNs, etc. are discussed in the companion paper \cite{BPEHS}.)

\begin{figure}
\begin{center}
\includegraphics[width=0.45\textwidth]{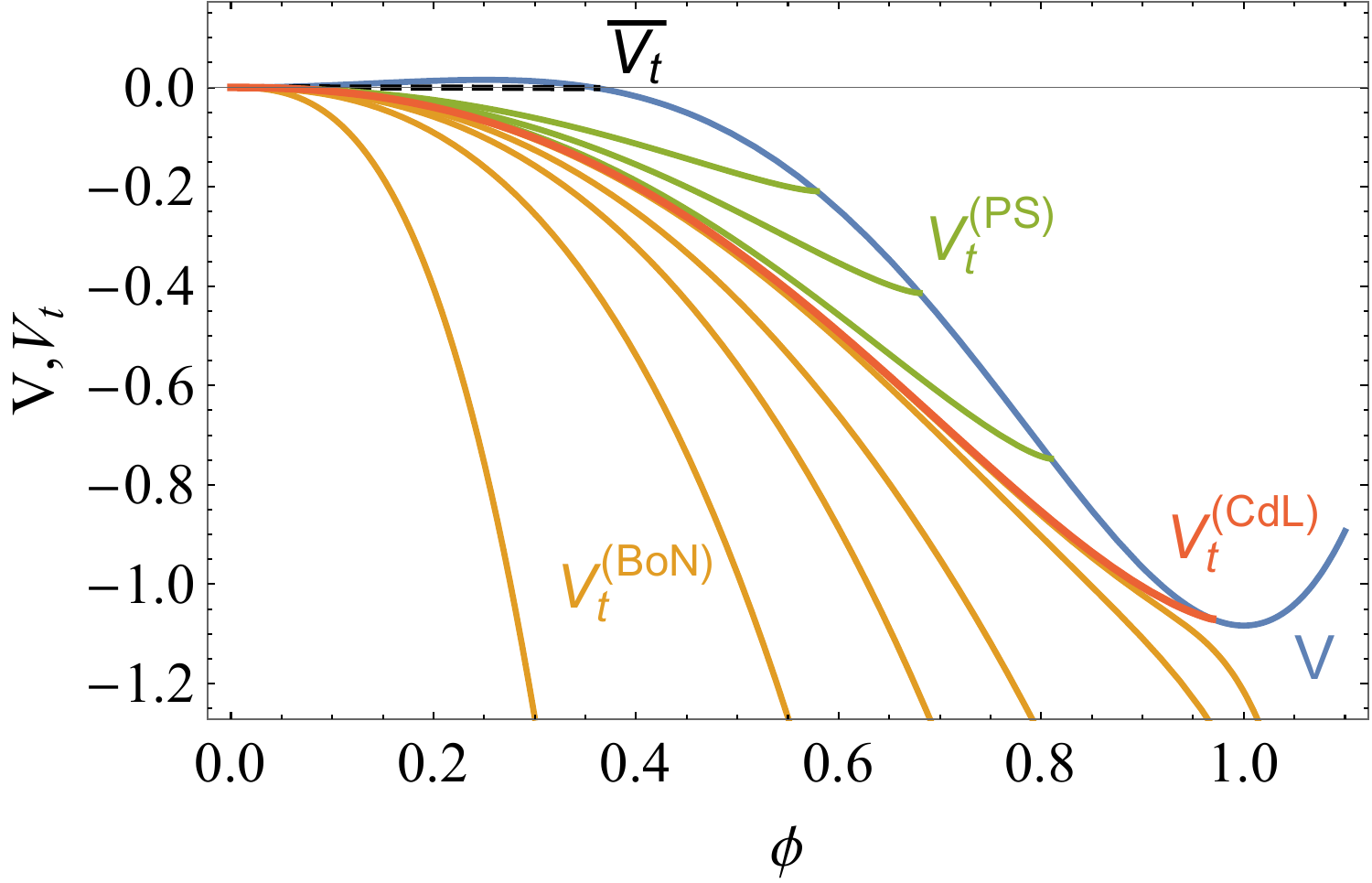}
\includegraphics[width=0.45\textwidth]{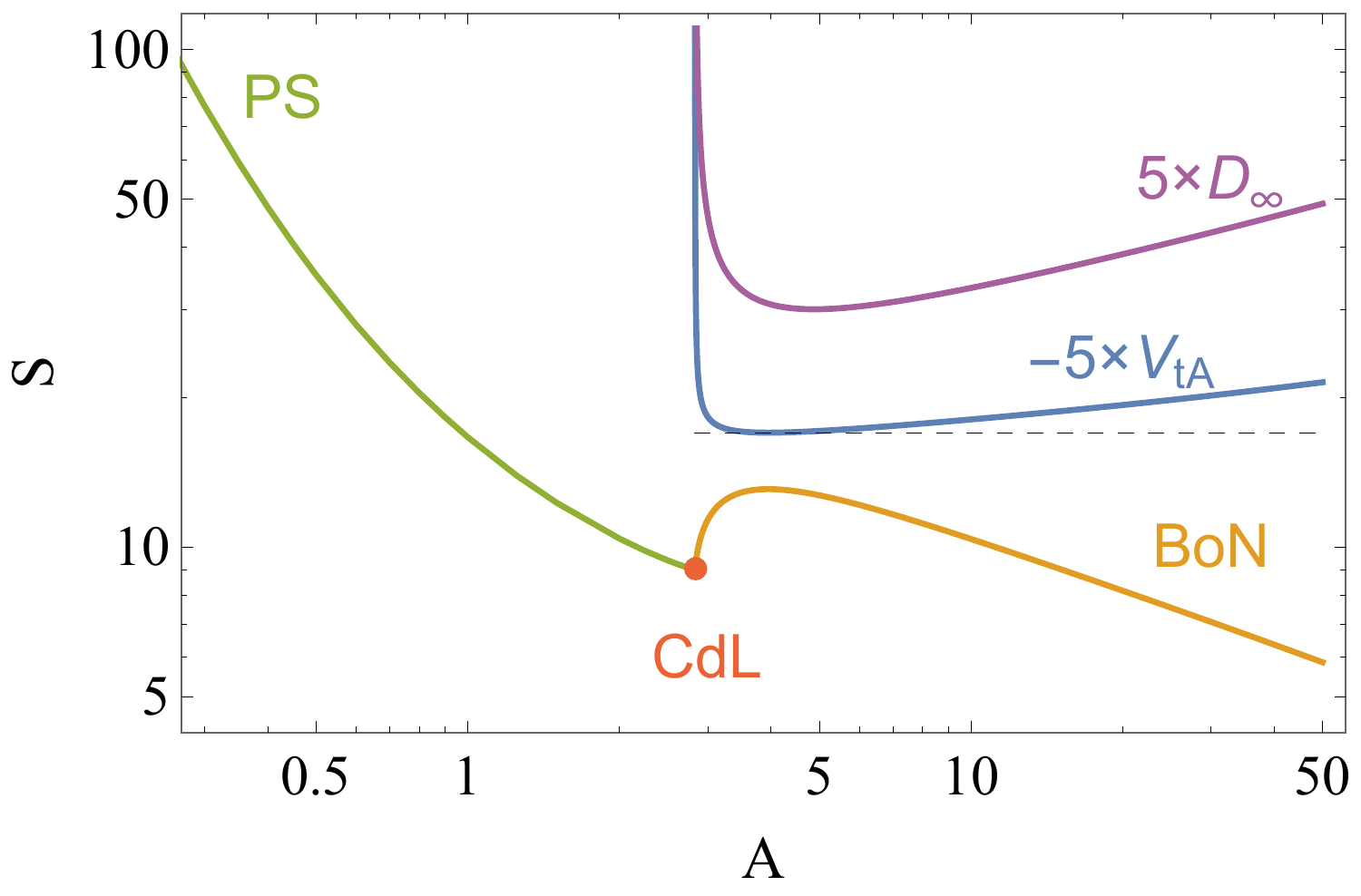}
\end{center}
\caption{
Top: Potential (\ref{Vcex}) with $V_\pp=0$ and tunneling potentials $V_t(A;\phi)$: $\overline{V_t}$ (black dashed); pseudo-bounces (green); CdL bounce (red) and BoNs (orange). 
Bottom: Tunneling action $S$ for the $V_t(A;\phi)$. 
For the BoN range of $A$, (rescaled) prefactors  $V_{tA}$ and  $D_\infty $. 
\label{fig:PSCdLBoN}
}
\end{figure}

The Minkowski case ($V_\pp=0$) is shown in  figure~\ref{fig:PSCdLBoN}. In the upper plot, the ($D=0$) $\overline{V_t}$ of (\ref{Vtc})
(black dashed line) touches the potential beyond the barrier, signaling a CdL  instability of the false vacuum. Below $\overline{V_t}$ we find $V_t(A;\phi)$ solutions, labeled by the parameter $A$ of the low field $V_t$ expansion: the CdL instanton solution (red line) for $A=A_{CdL}\simeq 2.8$; pseudo-bounce solutions (green lines) for $A<A_{CdL}$; and unbounded BoN solutions (orange lines) for $A>A_{CdL}$.

Figure~\ref{fig:PSCdLBoN}, bottom plot, gives the tunneling action of the $V_t$ solutions just described. The action of pseudo-bounces diverges at $A\to 0$ (when $V_t\to\overline{V_t}$) and, interestingly, the BoN action beyond the CdL point can be larger or smaller  than $S_{CdL}$.
Thus,  the BoN decay channel not always dominates.

The BoNs obtained  are of type 0, with $V_t\sim V_{tA}e^{\sqrt{6}\phi}$ and $D\sim D_\infty e^{\sqrt{8/3}\phi}$ as $\phi\to\infty$. 
The lower plot of figure~\ref{fig:PSCdLBoN}  shows $D_\infty(A)$ and $-V_{tA}(A)$ which is bounded below by $-V_{tA*} = -V_{tA}(A_*)$ (black dashed line). That minimum is reached  when $S$ is maximal. 
 
In a given $4+1$ theory, $V_{tA}$ is determined by  \eqref{VUVminus} (with $d=1$).
When the bound \eqref{eq:dyConst1} is satisfied,  there are two possible BoNs, corresponding to the two solutions of  $R_{KK}^2=\frac{3}{4 (-V_{tA}(A))}$. The solution with lowest tunneling action (thus the relevant one) lies in the branch of solutions extending from the action maximum to values below $S_{CdL}$ ($A>A^*$).

BoN decays are forbidden if $R_{KK}^2 > \frac{3}{4 (-V_{tA*})}$ (although CdL decay is still open). This dynamical quench with finite $S$ can happen even for a dS vacuum  \cite{BPEHS}, in contrast with the standard quenching of decay, which only occurs for $V_\pp\leq 0$. 
If we require the KK and 4d EFT scales to be well separated ($R_{KK}$ small compared to the typical EFT length-scale) this needs large $-V_{tA}$ due to \eqref{eq:dyConst1}. In this limit, where the EFT is well under control,  BoN decay is always  allowed, and becomes the fastest decay channel.

\section{Summary \label{sec:concl}}

The $V_t$ method greatly facilitates the study of which modulus potentials $V(\phi)$ admit BoN decays and which types of BoN exist. We identify four types of BoN, with different asymptotics  in the compactification limit ($\phi\to\infty$,  the BoN core), 
see table~\ref{table:types}. Type 0/$-$  BoNs can appear if the compact space is a $S^d$ sphere, while Type $+$ or $-^*$ BoNs need more complicated compact geometries, and/or the presence of some UV defect 
at the BoN core. 

For BoNs of types 0 or $-$, there are simple relations between the asymptotics of $V,V_t$ and the BoN  geometry in the $4+d$ theory (like the KK radius, $R_{KK}$).
Such relations tell which BoNs are relevant for a given theory.
For potentials not growing as fast as $e^{\sqrt{6\kappa}\phi}$, we find a continuous family of type 0 BoN solutions labeled by some parameter $p$, with 
$V_t(p;\phi)\simeq V_{tA}(p) e^{\sqrt{6}\phi}$ and $
D(p;\phi)\simeq D_\infty(p) e^{\sqrt{8/3}\phi}$ for $\phi\to\infty$. Fixing the compactification scale, $R_{KK}$, selects a finite number of BoNs from the family [each with different action]. 
The number of such selected BoNs is model dependent in the following way. 

When the modulus has a single vacuum (or if gravity forbids its decay) the BoN is unique (for fixed $R_{KK}$). If the vacuum is a Minkowski or AdS one, there is an upper critical limit $R_{KK}^*$ for which the BoN has infinite action and radius and turns into an end-of-the-world brane. For $R_{KK}>R_{KK}^*$, BoN decay is forbidden (CdL-like dynamical quenching). 

When the scalar potential has additional vacua and admits standard decay channels (CdL/HM) to them, there are (at least) two BoNs (the one with lowest action being the relevant one). In this case there is also a critical $R_{KK}^*$ which corresponds to the merging of the two BoN solutions into one with finite action. For $R_{KK}>R_{KK}^*$  BoN decay is again dynamically forbidden.

\vspace{-0.5cm}

\section*{Acknowledgments\label{sec:ack}} 

J.J.B.-P., J.R.E and J.H. are supported in part by PID2021-123703NB-C21, PID2022-142545NB-C22 and PID2021-123017NB-I00, funded by ``ERDF, A way of making Europe'' and by MCIN/ AEI/10.13039/501100011033. J.J.B.-P. is supported by the Basque Government grant (IT-1628-22) and the Basque Foundation for Science (IKERBASQUE). J.R.E. and J.H. are supported by  IFT Centro de Excelencia Severo Ochoa CEX2020-001007-S. J.H is supported by the FPU grant FPU20/01495 from the Spanish Ministry of Education and Universities.



\begin{thebibliography}{99}

\bibitem{Coleman}
S.R.~Coleman,
Phys. Rev. D \textbf{15} (1977) 2929
[erratum: Phys. Rev. D \textbf{16} (1977), 1248].

\bibitem{CdL}
S.R.~Coleman and F.~De Luccia,
Phys. Rev. D \textbf{21} (1980) 3305.

\bibitem{BoN}
E.~Witten,
Nucl. Phys. B \textbf{195} (1982) 481.

\bibitem{BS}
J.~J.~Blanco-Pillado and B.~Shlaer,
Phys. Rev. D \textbf{82} (2010) 086015
\arXiv{1002.4408}{th}.

\bibitem{Blanco-Pillado:2010vdp}
J.J.~Blanco-Pillado, H.S.~Ramadhan and B.~Shlaer,
JCAP \textbf{10} (2010) 029
\arXiv{1009.0753}{th}.

\bibitem{Blanco-Pillado:2016xvf}
J.J.~Blanco-Pillado, B.~Shlaer, K.~Sousa and J.~Urrestilla,
JCAP \textbf{10} (2016) 002
\arXiv{1606.03095}{th}.

\bibitem{Ooguri:2017njy}
H.~Ooguri and L.~Spodyneiko,
Phys. Rev. D \textbf{96} (2017) 026016
\arXiv{1703.03105}{th}.

\bibitem{Dibitetto:2020csn}
G.~Dibitetto, N.~Petri and M.~Schillo,
JHEP \textbf{08} (2020) 040
\arXiv{2002.01764}{th}.

\bibitem{CobConj}
J.~McNamara and C.~Vafa,
\arXiv{1909.10355}{th}.

\bibitem{DFG}
M.~Dine, P.J.~Fox and E.~Gorbatov,
JHEP \textbf{09} (2004) 037
\arXivold{th/0405190}.

\bibitem{DGL}
P.~Draper, I.G.~Garcia and B.~Lillard,
Phys. Rev. D \textbf{104} (2021) 12
\arXiv{2105.08068}{th};
JHEP \textbf{12} (2021) 154
\arXiv{2105.10507}{th}.

\bibitem{EEg}
J.R.Espinosa,
 JCAP{\bf 07} (2018) 36, \arXiv{1805.03680}{th};
  Phys.\ Rev.\ D {\bf 100} (2019)  104007
  \arXiv{1808.00420}{th}.
  

\bibitem{HM}
S.~W.~Hawking and I.~G.~Moss,
Phys. Lett. B \textbf{110} (1982) 35.


\bibitem{PS}
J.R.~Espinosa,
Phys. Rev. D \textbf{100} (2019) 105002
\arXiv{1908.01730}{th};
J.R.~Espinosa and J.~Huertas,
JCAP \textbf{12} (2021) 12, 029
\arXiv{2106.04541}{th}.

\bibitem{BPEHS}
J.J.~Blanco-Pillado, J.R.~Espinosa, J.~Huertas and K.~Sousa, to appear.

\bibitem{Horowitz:2007pr}
G.T.~Horowitz, J.~Orgera and J.~Polchinski,
Phys. Rev. D \textbf{77} (2008), 024004
\arXiv{0709.4262}{th}.

\bibitem{Bomans:2021ara}
P.~Bomans, D.~Cassani, G.~Dibitetto and N.~Petri,
SciPost Phys. \textbf{12} (2022) 3, 099
\arXiv{2110.08276}{th}.


\bibitem{Hebecker}
B.~Friedrich, A.~Hebecker and J.~Walcher,
\arXiv{2310.06021}{th}.



\bibitem{GMSV}
I.~Garc\'{\i}a Etxebarria, M.~Montero, K.~Sousa and I.~Valenzuela,
\arXiv{2005.06494}{th}.
 
\bibitem{DynCo}
R.~Angius, J.~Calder\'on-Infante, M.~Delgado, J.~Huertas and A.M.~Uranga,
JHEP \textbf{06} (2022) 142
\arXiv{2203.11240}{th}.  

\end{thebibliography}
\end{document}